\documentclass[aps,preprint, floatfix]{revtex4}

\usepackage{graphicx,color}
\usepackage{psfrag}
\usepackage{amssymb}
\usepackage{eso-pic}
\usepackage[dvips,letterpaper]{geometry} 
\usepackage{amsmath}

\begin{document}

%
%
%
%
\def\oti{{\otimes}}
\def\lb{ \left[ }
\def\rb{ \right]  }
\def\tilde{\widetilde}
\def\bar{\overline}
\def\hat{\widehat}
\def\*{\star}
\def\[{\left[}
\def\]{\right]}
\def\({\left(}		\def\BL{\Bigr(}
\def\){\right)}		\def\BR{\Bigr)}
	\def\BBL{\lb}
	\def\BBR{\rb}
%
%
\def\zb{{\bar{z} }}
\def\zbar{{\bar{z} }}
\def\frac#1#2{{#1 \over #2}}
\def\inv#1{{1 \over #1}}
\def\half{{1 \over 2}}
\def\d{\partial}
\def\der#1{{\partial \over \partial #1}}
\def\dd#1#2{{\partial #1 \over \partial #2}}
\def\vev#1{\langle #1 \rangle}
\def\ket#1{ | #1 \rangle}
\def\rvac{\hbox{$\vert 0\rangle$}}
\def\lvac{\hbox{$\langle 0 \vert $}}
\def\2pi{\hbox{$2\pi i$}}
\def\e#1{{\rm e}^{^{\textstyle #1}}}
\def\grad#1{\,\nabla\!_{{#1}}\,}
\def\dsl{\raise.15ex\hbox{/}\kern-.57em\partial}
\def\Dsl{\,\raise.15ex\hbox{/}\mkern-.13.5mu D}
%
%
\def\ga{\gamma}		\def\Ga{\Gamma}
\def\be{\beta}
\def\al{\alpha}
\def\ep{\epsilon}
\def\vep{\varepsilon}
\def\la{\lambda}	\def\La{\Lambda}
\def\de{\delta}		\def\De{\Delta}
\def\om{\omega}		\def\Om{\Omega}
\def\sig{\sigma}	\def\Sig{\Sigma}
\def\vphi{\varphi}

%
%
\def\CA{{\cal A}}	\def\CB{{\cal B}}	\def\CC{{\cal C}}
\def\CD{{\cal D}}	\def\CE{{\cal E}}	\def\CF{{\cal F}}
\def\CG{{\cal G}}	\def\CH{{\cal H}}	\def\CI{{\cal J}}
\def\CJ{{\cal J}}	\def\CK{{\cal K}}	\def\CL{{\cal L}}
\def\CM{{\cal M}}	\def\CN{{\cal N}}	\def\CO{{\cal O}}
\def\CP{{\cal P}}	\def\CQ{{\cal Q}}	\def\CR{{\cal R}}
\def\CS{{\cal S}}	\def\CT{{\cal T}}	\def\CU{{\cal U}}
\def\CV{{\cal V}}	\def\CW{{\cal W}}	\def\CX{{\cal X}}
\def\CY{{\cal Y}}	\def\CZ{{\cal Z}}

\def\rvac{\hbox{$\vert 0\rangle$}}
\def\lvac{\hbox{$\langle 0 \vert $}}
\def\comm#1#2{ \BBL\ #1\ ,\ #2 \BBR }
\def\2pi{\hbox{$2\pi i$}}
\def\e#1{{\rm e}^{^{\textstyle #1}}}
\def\grad#1{\,\nabla\!_{{#1}}\,}
\def\dsl{\raise.15ex\hbox{/}\kern-.57em\partial}
\def\Dsl{\,\raise.15ex\hbox{/}\mkern-.13.5mu D}
%
%
%
\font\numbers=cmss12
\font\upright=cmu10 scaled\magstep1
\def\stroke{\vrule height8pt width0.4pt depth-0.1pt}
\def\topfleck{\vrule height8pt width0.5pt depth-5.9pt}
\def\botfleck{\vrule height2pt width0.5pt depth0.1pt}
\def\Zmath{\vcenter{\hbox{\numbers\rlap{\rlap{Z}\kern
0.8pt\topfleck}\kern 2.2pt
                   \rlap Z\kern 6pt\botfleck\kern 1pt}}}
\def\Qmath{\vcenter{\hbox{\upright\rlap{\rlap{Q}\kern
                   3.8pt\stroke}\phantom{Q}}}}
\def\Nmath{\vcenter{\hbox{\upright\rlap{I}\kern 1.7pt N}}}
\def\Cmath{\vcenter{\hbox{\upright\rlap{\rlap{C}\kern
                   3.8pt\stroke}\phantom{C}}}}
\def\Rmath{\vcenter{\hbox{\upright\rlap{I}\kern 1.7pt R}}}
\def\Z{\ifmmode\Zmath\else$\Zmath$\fi}
\def\Q{\ifmmode\Qmath\else$\Qmath$\fi}
\def\N{\ifmmode\Nmath\else$\Nmath$\fi}
\def\C{\ifmmode\Cmath\else$\Cmath$\fi}
\def\R{\ifmmode\Rmath\else$\Rmath$\fi}

\def\barray{\begin{eqnarray}}
\def\earray{\end{eqnarray}}
\def\beq{\begin{equation}}
\def\eeq{\end{equation}}

\def\n{\noindent}

\def\Tr{\rm Tr} 
\def\xvec{{\bf x}}
\def\kvec{{\bf k}}
\def\kvecp{{\bf k'}}
\def\omk{\om{\kvec}} 
\def\dk#1{\frac{d\kvec_{#1}}{(2\pi)^d}}
\def\2pid{(2\pi)^d}
\def\ket#1{|#1 \rangle}
\def\bra#1{\langle #1 |}
\def\vol{V}
\def\adag{a^\dagger}
\def\rme{{\rm e}}
\def\Im{{\rm Im}}
\def\pvec{{\bf p}}
\def\fermiS{\CS_F}
\def\cdag{c^\dagger}
\def\adag{a^\dagger}
\def\bdag{b^\dagger}
\def\vvec{{\bf v}}
\def\muhat{{\hat{\mu}}}
\def\vac{|0\rangle}
\def\pcut{{\Lambda_c}}
\def\chidot{\dot{\chi}}
\def\gradvec{\vec{\nabla}}
\def\psitilde{\tilde{\Psi}}
\def\psibar{\bar{\psi}}
\def\psidag{\psi^\dagger} 
\def\m{m_*}
\def\up{\uparrow}
\def\down{\downarrow}
\def\Qo{Q^{0}}
\def\vbar{\bar{v}}
\def\ubar{\bar{u}}
\def\smallhalf{{\textstyle \inv{2}}}
\def\smallsqrt{{\textstyle \inv{\sqrt{2}}}}
\def\rvec{{\bf r}}
\def\avec{{\bf a}}
\def\pivec{{\vec{\pi}}}
\def\svec{\vec{s}} 
\def\phivec{\vec{\phi}}
\def\daggerc{{\dagger_c}}
\def\Gfour{G^{(4)}}
\def\dim#1{\lbrack\!\lbrack #1 \rbrack\! \rbrack }
\def\qhat{{\hat{q}}}
\def\ghat{{\hat{g}}}
\def\nvec{{\vec{n}}}
\def\bull{$\bullet$}
\def\ghato{{\hat{g}_0}}
\def\r{r}
\def\deltaq{\delta_q}
\def\gcharge{g_q}
\def\gspin{g_s}
\def\deltas{\delta_s}
\def\gQC{g_{AF}} 
\def\ghatqc{\ghat_{AF}}
\def\xqc{x_{AF}}
\def\mhat{\hat{m}}
\def\xup{x_2}
\def\xdown{x_1}
\def\sigmavec{\vec{\sigma}}
\def\xopt{x_{\rm opt}}
\def\Lambdac{{\Lambda_c}}
\def\angstrom{{{\scriptstyle \circ} \atop A}     }
\def\AA{\leavevmode\setbox0=\hbox{h}\dimen0=\ht0 \advance\dimen0 by-1ex\rlap{
\raise.67\dimen0\hbox{\char'27}}A}
\def\ratio{\gamma}
\def\Phivec{{\vec{\Phi}}}
\def\singlet{\chi^- \chi^+} 
\def\mhat{{\hat{m}}}

\def\Im{{\rm Im}}
\def\Re{{\rm Re}}

\def\xstar{x_*}

\def\sech{{\rm sech}}

\def\Li{{\rm Li}}

\def\dim#1{{\rm dim}[#1]}

\def\ep{\epsilon}

\def\free{\CF}

\def\Fhat{\digamma}

\def\ftilde{\tilde{f}}

\def\muphys{\mu_{\rm phys}}

\def\xitilde{\tilde{\xi}}

\def\CI{\mathcal{I}}

\def\nhat{\hat{n}}

\def\ef{\epsilon_F}

\def\pvec{{\bf p}}

\def\Kvec{{\bf K}}

\def\thatprime{{\hat{t}\,'}}

\title{Thermodynamics of the two-dimensional Hubbard model based on the exact 
two-body S-matrix}
\author{ Andr\'e  LeClair}
\affiliation{Newman Laboratory, Cornell University, Ithaca, NY} 
\affiliation{ Centro Brasileiro de Pesquisas F\'isicas,  Rio de Janeiro}  

\bigskip\bigskip\bigskip\bigskip

\begin{abstract}
 
A new  analytic treatment of the two-dimensional Hubbard model 
at finite temperature and chemical potential is presented.   A next nearest neighbor hopping term of
strength $t'$ is included.    This analysis is  based upon 
a    formulation of  the statistical mechanics of particles  in terms of 
the S-matrix.      We show that for $U/t$ large enough, 
a region of attractive interactions exists near the Fermi surface  due to multi-loop quantum 
corrections.   For $t'=-0.3$, these  attractive interactions exist for $U/t >  6.4$.    
Our analysis suggests that superconductivity may not exist  for $t'=0$.   
Based on the existence of solutions of the integral equation for the pseudo-energy,  
we provide evidence for a phase transition  
and estimate $T_c /t \approx 0.02$ for $U/t=7.5$ and $ t' /t =-0.3$ at hole doping $0.15$.

\end{abstract}

\maketitle

\section{Introduction}

The Hubbard model in two spatial dimensions plays a central role in the modern theory
of strongly correlated electrons.   Since it is believed to be a good microscopic model for
the underlying physics of high $T_c$ superconductivity (HTSC)\cite{Anderson},   it has been studied
extensively over the past two decades. 
A partial list of publications on the thermodynamics of the Hubbard model is
\cite{Hone,Hirsch,Dagotto,Georges,Dongen,Potthoff,Bonca,Li,MonteCarlo}. 
   Despite this effort,   many of its important 
properties are not currently well-understood,  and  it remains to be established definitively 
 that it possesses
all of the main features of HTSC.    For instance,  the precise mechanism  
that leads to attractive d-wave pairing,  which must arise from  
the purely electronic Coulomb repulsion,   is still not well understood.    Analytic methods to date 
are rather limited since the model is in the  strong coupling regime.    
Lattice Monte-Carlo methods on the other hand are limited to small lattices and suffer
from the fermion sign problem,  especially at non-zero doping and sufficiently low 
temperatures.    
For these reasons,   any new analytic methods,   though approximate,   may shed new light
on the problem.    

In this work we present an  analytic approach to the thermodynamics of particles at finite density
and temperature  based on the reformulation 
of the statistical mechanics of particles in terms of the {\it zero temperature and density}  particle-particle 
  S-matrix developed in 
\cite{LeClairS,PyeTon}.     As explained there,  the potential advantage of this method is
that,  unlike the usual diagramatic  Matsubara approach to finite temperature field theory, 
it disentangles the  zero temperature dynamics from the quantum statistical sums.   
Some remarks clarifying the nature of this formalism,  and the approximations made, 
are called for.   
The approach was modeled after the thermodynamic Bethe ansatz\cite{YangYang},  
 which is exact  for integrable models in 
one spatial dimension since the N-body S-matrix factorizes into 2-body S-matrices,  
Our method indeed reduces to the thermodynamic Bethe ansatz in the two-body scattering approximation,
as shown in \cite{PyeTon}  for the interacting 1d Bose gas.    The main approximation we make is
that we consider only many-body processes that involve arbitrary numbers of primitive binary collisions.    
I.e. we neglect processes that in some sense involve 3 or more particles colliding simultaneously,  which
is not the same as ignoring the many-body aspect of the problem altogether.    
In this non-relativistic context,  it is well-known that the two-body S-matrix can be calculated
exactly,  thus in some regards the method is non-perturbative.     Although   this is not a fully controlled
approximation,   it has been demonstrated to give reliable results for other strong-coupling problems,
in particular 
the critical point of the 2d bosonic gas and more importantly  to Bose and Fermi gases in the 
3d scale-invariant  unitary limit\cite{PyeTon,PyeTon2}.      For instance for the unitary Fermi gas on 
the BEC side of the crossover,    the critical temperature calculated
is consistent with Monte-Carlo methods,  and the ratio of the viscosity to entropy density agrees very
well  with the most recent experimental data\cite{viscosity}.    
It should be pointed out that the exact Bethe-ansatz solution of the 1d Hubbard model 
exhibits two additional holon excitations\cite{Essler},  which can be inferred from poles in the S-matrix
of the fundamental fermions;   we have no evidence for such excitations in the 2-dimensional case,
thus it is unclear whether a meaningful comparison with the 1d case can be made.

Our conventions for the Hubbard model are described in the next section.    We include
a next nearest neighbor hopping term of strength $t'$,  since it is known to be non-zero
in the cuprates;  as we will show,  its effects are important.   In section III,  the effective momentum dependent coupling,  which is the  kernel $G(\kvec_1, 
\kvec_2 )$  constructed from  the logarithm of the 2-body S-matrix  in the
integral equation satisfied by the pseudo-energy,  is  analyzed.   We show that   there 
exists a  band of
attractive interactions near the half-filled Fermi surface  for $U/t $ large enough, 
 and $t'$ plays an important role in determining this property.  
We emphasize that no approximations are   made in section III,  since,  as stated above,  the 
2-body S-matrix can be calculated exactly,    and this attractive region
exists regardless of the subsequent approximations we make in studying the thermodynamics.   
This attractive mechanism, which arises from quantum loop corrections, 
 appears to be different than other mechanisms discussed in this  context,
such as those based on spin fluctuations or the resonating valence bond picture.   
It is thus important to investigate the consequences of these attractive interactions and how
they might be connected to HTSC,  and this paper is a first step in this direction.     
      In section IV  the
 S-matrix based formalism we utilize for calculating thermodynamic properties is 
 reviewed and specialized to the Hubbard gas.    In section V  the free energy is analyzed,
 and we present  some  evidence for phase transitions.

\section{Hubbard model conventions}

The Hubbard model describes fermionic particles  with spin,   hopping between the sites
of a square lattice, subject to strong local coulombic repulsion.  
The lattice hamiltonian is 
\beq
\label{Hublat}
H= - t \sum_{<i,j>, \alpha = \up ,\down} \(  c^\dagger_{\rvec_i , \alpha} 
c_{\rvec_j , \alpha}   \)
 - t' \sum_{<i,j>', \alpha = \up ,\down} \(  c^\dagger_{\rvec_i , \alpha} 
c_{\rvec_j , \alpha}   \) 
 +  U \sum_{\rvec} n_{\rvec\up} n_{\rvec \down} 
\eeq
where $\rvec_{i,j} ,  \rvec$ are sites of the lattice,  $<i,j>$ denotes
nearest neighbors,  $n = c^\dagger c $ are densities,   and $c^\dagger, c$ satisfy canonical anti-commutation
relations.    
For both  cuprates  LSCO and BSCO,  $U/t \approx 13$.   
We have also included a next to nearest neighbor hopping term $t'$, 
since it is not difficult to incorporate  into the formalism,  and it is known to be  non-zero
for high $T_c $ materials.     As we will see,  it can play a significant role.  
For  LSCO and BSCO,   $t'/t$   approximately equals $-0.1$ and
$-0.3$ respectively;   in our analysis  below we set $t'/t = -0.3$.

We introduce the two fields $\psi_{\up, \down}$ 
and the action
\beq
\label{action}
S = \int d^2 \rvec \,  dt \( \sum_{\alpha= \up, \down}  i  \,  \psi^\dagger_\alpha \d_t \psi_\alpha -
\CH \)
\eeq
where $\CH$ is the hamiltonian density.  
The field has the following expansion characteristic of a non-relativistic theory since it only involves
annihilation operators,
\beq
\label{field}
\psi_\alpha (\rvec ) = \int \frac{d^2 \kvec}{2\pi} ~ 
c_{\kvec, \alpha}  \, e^{i \kvec \cdot \rvec} 
\eeq
and satisfies 
\beq
\label{concom}
\{ \psi_\alpha (\rvec ) , \psi^\dagger_{\alpha'} (\rvec') \} = 
\delta(\rvec - \rvec') \delta_{\alpha, \alpha'}
\eeq
Since we have represented sums over lattice sites $\rvec$ as $\int  d^2 \rvec /a^2$,  
where $a$ is the lattice spacing,  
$c_\rvec  =  a  \psi (\rvec )$.   
The  free part of the hamiltonian is then 
\beq
\label{freeH}
H_{\rm free} =  \int d^2 \kvec ~  \omega_\kvec \sum_\alpha c^\dagger_{\kvec, 
\alpha} c_{\kvec, \alpha} 
\eeq
with  the 1-particle energy
\beq
\label{omeg}
\omega_\kvec =  - 2 t \(  \cos( k_x a)  + \cos (k_y a ) \)   -4 t' \cos( k_x  a)  \cos (  k_y  a)
\eeq
where  $t$ taken to be   positive. 
   In the sequel it is implicit
that $\kvec$ is restricted to the first Brillouin zone, 
$-\pi/a \leq k_{x,y} \leq \pi/a$   

The interaction part of the hamiltonian is local,  and becomes 
 a continuum integral:      
\beq
\label{hint} 
H_{\rm int} =  \frac{u}{2}   \int d^2 \rvec ~~  \psi^\dagger_\up \psi_\up 
\psi^\dagger_\down \psi_\down 
\eeq
where $u = 2  U a^2$.   The model is now viewed  as a quantum
fermionic gas,  where the only effect of the lattice is in the free particle energies
$\omega_\kvec$.        
 
  The field $\psi$ has dimensions of  inverse length,
and the coupling $u$ units of ${\rm energy}  \cdot {\rm length}^2 $.   
In the sequel we will scale out the dependence on $t$ and the lattice spacing $a$,  
and physical quantities will then depend on the dimensionless coupling
\beq
\label{gdef} 
g =   \frac{u}{a^2 t}   = \frac{2U}{t}
\eeq
Positive $g$ corresponds to
repulsive interactions.

\section{The effective momentum dependent coupling and the possible  origin of attractive interactions}

In the finite temperature formalism developed in \cite{LeClairS,PyeTon},    the occupation numbers
$f$ are parameterized in terms of a pseudo-energy $\vep(\kvec)$ in the same manner as
for a  free theory:   $f = 1/(e^{\beta \vep} +1 )$.    In the approximation that only  the many-body processes 
built out of primitive 2-body collisions are retained,   the pseudo-energy satisfies an integral equation 
based on a kernel $G(\kvec_1 , \kvec_2)$  which is related to the logarithm of the 
2-particle  S-matrix.    This approach to the  thermodynamics  will be reviewed 
in the next section.   The final result derived in \cite{PyeTon} involves
only the particle-particle S-matrix at zero temperature and density,  which can be calculated exactly.
The temperature and density enter the formalism in the integral equation for the pseudo-energy, 
thus this formalism does not require particle-particle or particle-hole Green's functions at finite
temperature and chemical potential.    
In this section we  study the main features of the kernel  and  demonstrate that  there are  regions
of the Brillouin zone where the interactions are effectively attractive,  even though the bare model 
has repulsive interactions.          We we wish to emphasize that no approximations are made in
obtaining the results presented in this section,  which essentially amount to  quantum corrections to 
scattering,  and some conclusions  are independent of the thermodynamics studied  in subsequent 
sections.

\subsection{Structure of the kernel}

As described in \cite{PyeTon},  
the kernel has the following structure:
\beq
\label{Gstructure}
G(\kvec_1 , \kvec_2 )   =  - \frac{i}{2 \CI} \log \( 1 + 2 i \CI \CM \) 
\eeq
where $\CM$ is the  2-body  scattering amplitude. 
(We are suppressing the momentum dependence.) 
  $\CI$ represents
the available phase space for two-body scattering:
\beq
\label{cI}
\CI =  \inv{4\pi} \int d^2 \pvec \, \delta(E - \omega_\pvec - \omega_{\Kvec
- \pvec} )
\eeq
where $E$ and $\Kvec$ are the total energy and momentum of the two
incoming particles with momenta $\kvec_1$ and $\kvec_2$:
\beq
\label{total}
E = \omega_{\kvec_1} + \omega_{\kvec_2},  ~~~~~
\Kvec = \kvec_1 + \kvec_2
\eeq

All energy and temperature scales, $E$, $T$, the chemical potential $\mu$,  and $t'$, will be expressed in units of the hopping parameter $t$. 
We thus scale $t$ out of $\omega_\kvec$ so that henceforth  $\omega_\kvec $ equals 
(\ref{omeg}) divided by $t$.     We will also rescale $\kvec$ by $1/a$ so that 
$ -\pi  \leq  k_{x,y} \leq \pi$.     Since  $\CI$ is proportional to $1/t$, 
$G \propto t$,   and henceforth $G$ will represent  $G/t$,  which is dimensionless after scaling
out factors of $a$ also.      The kernel $G$   then depends only on the dimensionless coupling 
$g $  defined in  eq. (\ref{gdef}),  and  the momenta.

The scattering amplitude can be computed by summing multi-loop 
ladder diagrams\cite{PyeTon},  which factorize into 1-loop integrals in this non-relativistic 
context.   This  leads to 
\beq
\label{curlyM}
\CM =  \frac{ - g/2 }{1+ i g L /2} 
\eeq
where $g$ is the coupling defined in eq. (\ref{gdef}).  $L$ is a 
1-loop integral:
\barray
\nonumber
L &=&  \int_{-\infty}^\infty \frac{d \omega}{2\pi} \int \frac{d^2 p}{(2\pi)^2}
\( \frac{i}{\omega - \omega_\pvec + i \ep } \) 
\(\frac{i}{E   - \omega - \omega_{\Kvec - \pvec} + i \ep } \)
\\ 
\label{Loop} 
&=& i \int \frac{d^2 \pvec }{(2\pi)^2} ~  \inv{E  - \omega_\pvec - \omega_{\Kvec
- \pvec}  + 2 i \ep }
\earray 
where $\ep$ is small and positive.  
In the numerical analysis below  we set $\ep = 0.001$.  
Using $\Im (x + i \ep )^{-1}  = - \pi \delta(x)$,  one sees that 
\beq
\label{Lrealim}
L  =   \CI + i \gamma
\eeq
where $\CI$ is the phase space factor  in eq. (\ref{cI}) 
and is real and positive,  and $\gamma$ is  its imaginary part.  
Putting all of this together one has
\beq
\label{finalG}
G  =  - \frac{i}{2\CI} \log \(  \frac{1/g_R  - i \CI/2}{1/g_R + i \CI/2 } \)
\eeq
The imaginary part of the loop integral renormalizes the coupling $g$ 
\beq
\label{gR}
g_R  =  \frac{g}{1 - g \gamma/2} 
\eeq
Note that the manner in which $g_R$ enters the kernel leads to a  well-defined  large coupling
limit;      this was exploited for unitary quantum gases in \cite{PyeTon2} where
$g_R$ is proportional to the scattering length which goes to $\pm \infty$ in the unitary limit.

\begin{figure}[htb] 
\begin{center}
\hspace{-15mm} 
\includegraphics[width=13cm]{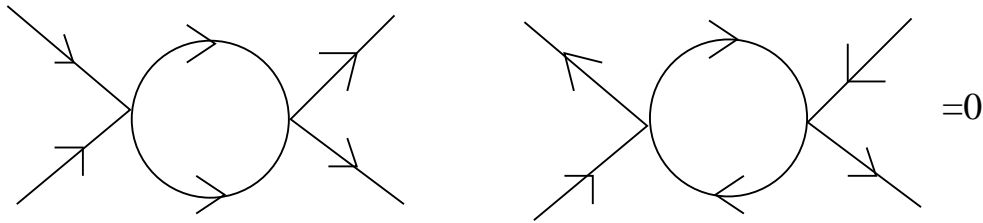} 
\end{center}
\caption{One-loop contributions to the S-matrix.  Only the diagram to the left is non-zero. }  
\vspace{-2mm}
\label{oneloops} 
\end{figure}

The argument
of the $\log$  in eq. (\ref{finalG})   can be identified as the   2-body S-matrix, which is unitary, 
i.e. $S^* S = 1$.    It should be emphasized that this is the {\it exact} two-body
S-matrix,  and this is possible because the model is non-relativistic.   More specifically,
the fields $\psi_{\up, \down}$ only involve annihilation operators,  in contrast to relativistic theories
which are expanded in both creation and annihilation operators.   
Let us elaborate on this important point,  which is well-known in other contexts,  such as 
non-relativistic quantum gases,  but often not completely explained.    
Represent the interaction vertex with two  incoming arrows for the annihilation operator fields
$\psi_{\up, \down}$ and two outgoing arrows for the creation fields $\psi^\dagger_{\up, \down}$.    
  Consider  for simplicity  the 1-loop contributions.  
Diagrams with a closed loop,  i.e. with arrows circulating in the same directions,  
 such as the second diagram in  Figure \ref{oneloops},  are zero because the integration 
over energy $\omega$ inside the loop has poles in the integrand that are either both in the upper
or lower half-plane,  so that the contour can be closed at infinity without picking up residues. 
(This is only true because our  formalism only involves the particle-particle S-matrix at zero 
temperature and density.)      In other 
words,  there is no ``crossing-symmetry''  as in relativistic theories,  where there are three  non-zero
 1-loop 
diagrams,  with different momentum dependence,   which are crossed versions of each other.   
  The non-zero multi-loop diagrams are 
only of the ``ladder type'',  which factorize,  and this was  implicitly used in \cite{PyeTon}.   
There is  actually no fermionic minus sign associated with each loop since the arrows do not
 form a {\it closed } loop. 
Since this S-matrix is exact to all orders in $g$,  
the thermodynamic formalism we will use embodies some non-perturbative
aspects of the problem,  although it still represents an approximation 
to the thermodynamics,  as explained in more detail in the next section.

\subsection{Origin of attractive interactions}

\def\thatprime{t'} 

The kernel $G$ by construction is real.    For small coupling $g$, 
$G$ is independent of momentum and equal to $-g/2$.  Thus $G$ may be viewed
as an effective,  momentum-dependent  coupling constant,  and provides valuable
information on the effective 2-body interactions at zero temperature.   When  
$G$ is negative the interactions are effectively repulsive,  otherwise they
are attractive.    The important point is that the renormalization of
$g$ to $g_R$  can in fact change the sign of $g_R$, which 
 changes the sign of $G$ due to the branch cut in the logarithm.    To demonstrate 
how this can happen,  we first perform the $p_y$ integral 
in $L$.   The result is 
\beq
\label{LCX}
L = \frac{i}{2 \pi^2} \int_{-\pi}^\pi dp_x  ~  
\inv{\sqrt{B^2 + C^2 - D^2} }   
\[  \log \( \frac{ (C-D) }{\sqrt{B^2 + C^2 - D^2}} \)   
- \log \(  \frac{(D-C)}{\sqrt{B^2 + C^2 - D^2 }} \) \]
\eeq
where 
\barray
\nonumber
B &=& 2 \sin K_y +  4 \thatprime \cos (K_x - p_x) \sin K_y  
\\ 
\label{BCD} 
C &=& 2+ 2 \cos K_y  + 4 \thatprime \cos p_x  + 4 \thatprime \cos (K_x - p_x)  \cos K_y 
\\
\nonumber
D &=&  E  + 2 \cos p_x + 2\cos(K_x - p_x)  + 2 i \ep 
\earray
(Recall $t'$ represents $t'/t$.)  
The sum  of the logarithms in the above formula is  simply $\log (-1) 
= \pm i \pi$;  however expressing the integral in this fashion ensures one
is on the proper branch.   The above formula proved to be
very useful for the numerical evaluation of the kernel.

From the expression  for the renormalized coupling $g_R$,  eq. (\ref{gR}),  one sees
that $g_R$ can become negative if $\gamma$ is positive  and $g$ large enough, 
$g> 2/\gamma$.    Let $k_1, k_2$  be in the center of mass frame, 
$\kvec_1= - \kvec_2 = \kvec $, so that $\Kvec = 0$,  and $L$ only depends on the total energy $E$.  
    The loop integral (\ref{LCX}) can be expressed in terms
of elliptic functions.    To regulate the integral,  we let   the  upper limit of the $p_x$ integral be $\pi - \kappa$,  and then
let $\kappa \to 0$.     The loop integral $L  \propto F(i \log (4 a/\kappa), b)$ where $F$ is the elliptic integral of the
first kind,  with $a = \sqrt{ (E-8+8t')/(E-8t')} $  and  $b = (E-8 t')^2 /((E+8t')^2 - 64) $.    As $\kappa \to 0$,  
we use  $\lim_{x\to \infty}  F(ix, b) =  i  K (1-b)$,  where $K$ is the complete elliptic integral of the first kind.    
The final result is 
\beq
\label{Lana}
L_{\Kvec =0} =  \frac{2}{\pi} \(   \frac{E-8t'}{(E+8+8t') (E(8-E) + 64 t' (t' -1) )} \)^{1/2}  \,
K\( \frac{32(Et' -2)}{(E+ 8 t')^2 -64}  \)
\eeq
with $E \to E + 2 i \epsilon$.      The flip in sign is a result of the
combination of logarithms  in eq. (\ref{LCX}).   

\begin{figure}[htb] 
\begin{center}
\hspace{-15mm} 
\psfrag{x}{$E/t$}
\psfrag{y}{$\gamma$}
\includegraphics[width=10cm]{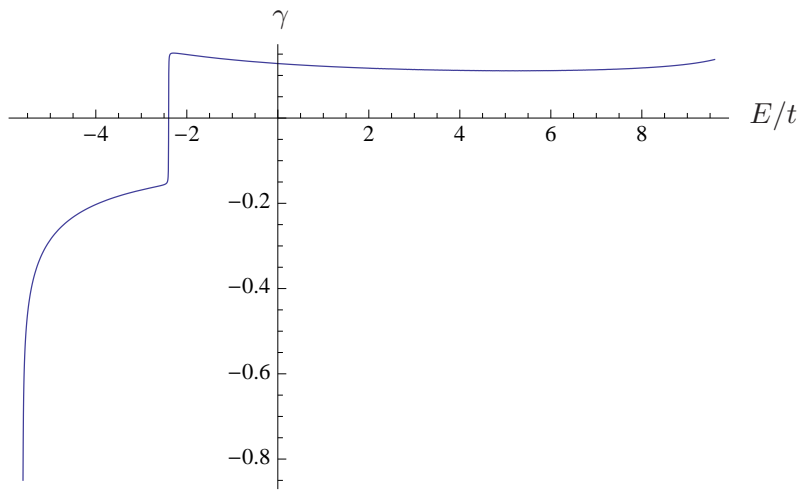} 
\end{center}
\caption{The imaginary part of the loop integral $\gamma$ as a function of energy $E$ for $t' /t =-0.3$.   }  
\vspace{-2mm}
\label{gamma} 
\end{figure}

The imaginary part of $L$, i.e. $\gamma$,  is plotted in Figure \ref{gamma}  for $t'=-0.3$.   
Since $\gamma$ is positive for large enough $E$,  one reaches the remarkable conclusion that for
$g$ large enough,  the effective interactions  can become attractive.   One can estimate 
this threshold for $g$  as $2/\gamma_{\rm max}$,  where $\gamma_{\rm max}$  is the
maximum value of $\gamma$ which occurs where $\gamma$ flips sign.    This value of $\gamma_{\rm max}$ can be obtained using the formula  (\ref{Lana}), and  is $\gamma_{\rm max} \approx 0.156$ 
for $t' =-0.3$.  This minimal threshold in $g$ should be contrasted with the Cooper instability,
which leads to superconductivity for arbitarily weak coupling.    This fact is a consequence of 
a logarithmic divergence in the analogous loop integrals,  which leads a gap proportional to 
$e^{-1/u} $  for some appropriate coupling $u$.   In the present context,  the  attractive interactions
arising from the above change of sign do not involve an analogous logarithmic divergence,  hence
the threshold.   
   
In summary,  we have shown that there are effectively attractive interactions above a threshold in
$U/t$,  e.g. for $t'=-0.3$,  attractive interactions exist for 
 $g>12.8$, i.e. $U/t > 6.4$.     
We repeated this  analysis for other values of $t'$ using the formula (\ref{Lana}), 
and our results for the mininum value of $g$ 
necessary for attractive interactions are shown in the table below. 
A minimal threshold for superconductivity was proposed in \cite{Kondo}, 
with minimal values of $U/t$ in the comparable range of $4-7$.   
On the other hand, the study in\cite{Raghu} indicates no threshold, namely, 
for the particular mechanism they they study, superconductivity exists for arbitrarily
low $U/t$.    This does not necessarily contradict our result,  since the attractive mechanism
we study here is essentially different, and we have not made a case yet that it is the one  responsible 
for superconductivity.

\begin{center}
\begin{tabular}{|c|c|}
\hline\hline
$t'/t$     &   $g_{\min}$   \\
\hline\hline 
$-0.1$  & $ 16.7   $             \\
$-0.2$  &  $ 14.3     $           \\
$-0.3$   & $ 12.8  $              \\
$-0.4 $    &  $10.0$    \\
\hline\hline 
\end{tabular}
\label{table1}
\end{center}

The change in sign of the effective coupling described above  is 
 reminiscent of  what is encountered 
in the BEC/BCS crossover of the non-relativistic continuum  three-dimensional unitary gas.  
 The model is defined with repulsive interactions, i.e.
positive coupling,  
however there is  a fixed point at  a 
negative coupling   $g_*$, independent of momenta.
     The scattering length is proportional to the renormalized coupling
$g_R = g/(1 - g/g_*)$.  
Just above the fixed point,  the scattering length goes to 
$-\infty$,  whereas just below it goes to $+\infty$.    The kernel $G$ for 
this model flips sign as one crosses the fixed point for the same reasons as above,
i.e. because  of  the branch-cut of the  logarithm\cite{PyeTon2}.     
Thus,   the effective interactions can be repulsive or attractive,  depending on which side of the fixed point one  sits,   even though the 
bare model  defined by the hamiltonian  had  only repulsive interactions.   
The main difference in the Hubbard gas is that the renormalized coupling depends on the 
momenta,   so that the interactions may become attractive in distinct regions of the 
Brillouin zone.  

 We mention that   a 
 change in sign of certain couplings under renormalization group flow 
 was found for a 2-chain Hubbard model  (2-legged ladder) in \cite{Balents}.    Such ladders are effectively
 1 dimensional,  and were mapped  onto an anisotropic Gross-Neveu model,  i.e. free Dirac
 fermions with marginal current-current interactions.      Since these Gross-Neveu models  are very 
 different from  those considered here,   
  it seems unlikely that the change of sign described in \cite{Balents}  is related to 
 the one described here,  since our model is intrinsically 2-dimensional,  and  the phenomenon 
 does not involve any renormalization group flow.

\begin{figure}[htb] 
\begin{center}
\hspace{-15mm} 
\psfrag{x}{$E$}
\psfrag{y}{$G$}
\psfrag{a}{$g=5$}
\psfrag{b}{$g=13.5, 14.$}
\psfrag{c}{$g=15$}
\psfrag{d}{$g = 20$}
\includegraphics[width=10cm]{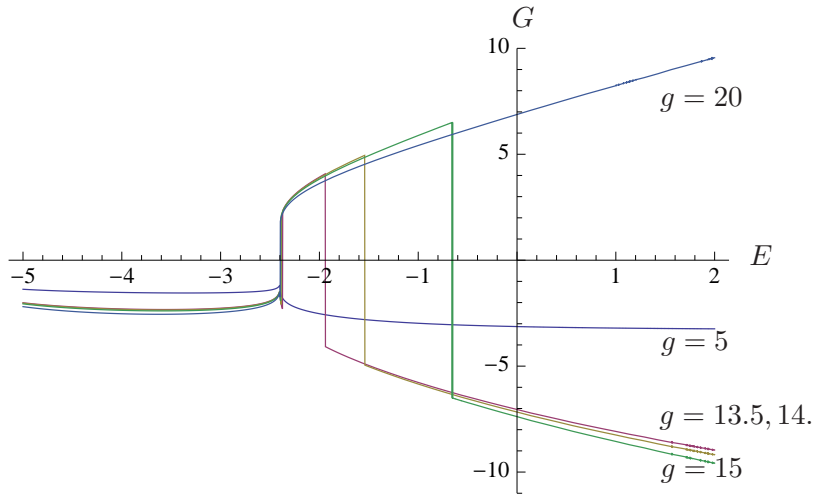} 
\end{center}
\caption{The kernel $G$  as a function of total energy $E$ for 
$g =5, 13.5, 14, 15, 20$,  and $t'/t = -0.3$  (Color figures on-line.)}  
\vspace{-2mm}
\label{GE520} 
\end{figure}

We now study the kernel $G(\kvec_1 , \kvec_2 )$  for $\kvec_1 = - \kvec_2 = \kvec$ 
and verify the above results.    
For these momenta,  the  kernel is only a function of the total energy $E$.   
In Figure \ref{GE520}, we plot $G(E)$ for the  values of the coupling
$g=5, 13.5,  14, 15, 20$  and $t' = -0.3$.     One observes that for  the smaller $g =5$,  the kernel is
everywhere negative. 
  One can verify  that for $g$ large enough, 
 the main features  do not depend strongly on $g$.    
 The most interesting region of $g$ is around $g\approx 13-15$ for $t'=-0.3$.   
 Comparing $g=14$ and $15$,  one sees that
for $g=14$ the attractive band is narrower,  and for $g=13$ essentially disappears.  
We will fix $g=15$ in our subsequent thermodynamic analysis,  since this is in the interesting region
and the attractive band is not too narrow.

\begin{figure}[htb] 
\begin{center}
\hspace{-15mm} 
\psfrag{x}{$E/t$}
\psfrag{y}{$G$}
\psfrag{low}{$t'=0$}
\psfrag{high}{$t' = -0.4$}
\includegraphics[width=10cm]{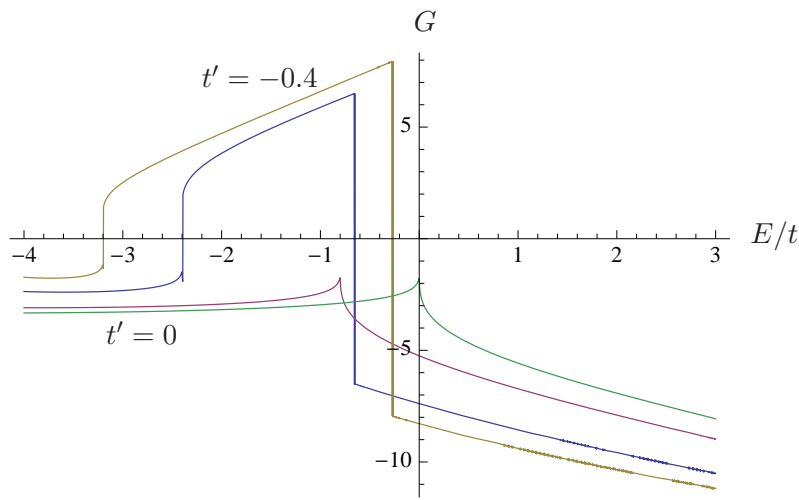} 
\end{center}
\caption{The kernel $G$  as a function of total energy $E$ for 
$g=15$   and $t'  = 0, -0.1, -0.3, -0.4$.}  
\vspace{-2mm}
\label{GE15} 
\end{figure}

Figure \ref{GE15}  shows the kernel 
for $g=15$ for $t'/t=0,-0.1, -0.3, -0.4$.  
One sees that  the interactions are effectively attractive only around a small region 
centered at $E\approx -1.5$ for $t'/t=-0.3$.  
An important feature of our analysis is that it clearly shows the importance of a non-zero $t'$.
For { \it fixed}  $g$,  if $| t' |$ is too small,   there is no attractive region,  as is apparent in Figure \ref{GE15}.  
If superconductivity  indeed arises from these attractive interactions,  
then this suggests  that superconductivity may not exist if $t'=0$.  
  There is actually some experimental evidence for this,  in that $T_c$ as a function of
$t'/t$ appears to extrapolate to zero\cite{tprimedata} .  
It should be pointed out however that  as one lowers $t'/t$,  attractive regions continue to exist as long
as one raises the coupling $g$,   as is evident in the table above for $g_{\rm min}$ as a function 
of $t'$.

\begin{figure}[htb] 
\begin{center}
\hspace{-15mm} 
\includegraphics[width=10cm]{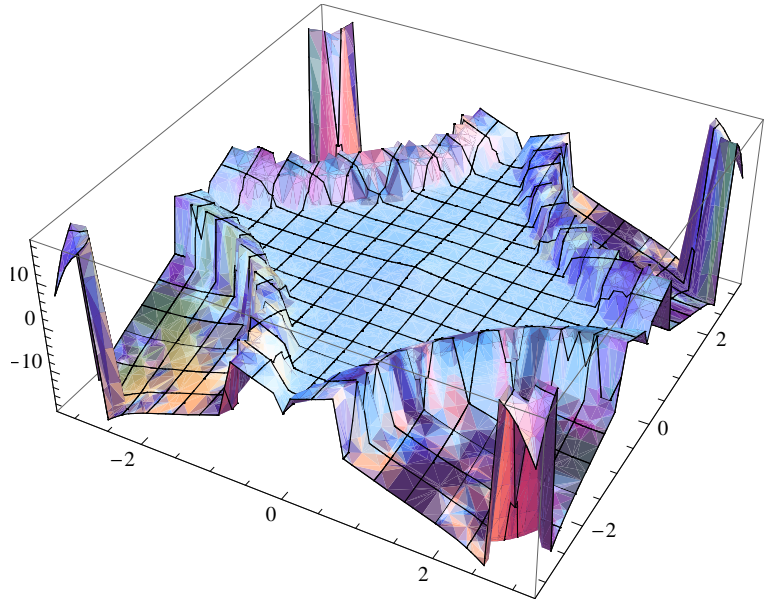} 
\end{center}
\caption{The kernel $G$ for Cooper pairs   in the first Brillouin zone for $g=15$
and $t'/t = -0.3$.
The horizontal axes are $  -\pi < k_{x,y} < \pi$ and the vertical 
axis is the effective coupling $G$.}  
\vspace{-2mm}
\label{G3D15} 
\end{figure}

It is also intructive to plot $G$ for pairs of opposite momentum  as a function of $k_x , k_y$
in the first Brillouin zone.  This is shown in Figure  \ref{G3D15}.
Again this shows  that the interactions are attractive 
in a narrow region around  half-filling.  The positive regions at the corners of the Brillouin zone are due to a divergence in the loop integral
which should be regularized;   however since we will be studying hole doping of the half-filled state,  
the densities will be low enough to be far from these regions,  so this regularization will be unnecessary. 
  Finally,  note that for low enough $E$,  
the interactions are always repulsive,   which    should imply    that at high enough
 doping the theory   should be  well-approximated by a Fermi liquid.

\begin{figure}[htb] 
\begin{center}
\hspace{-15mm} 
\includegraphics[width=12cm]{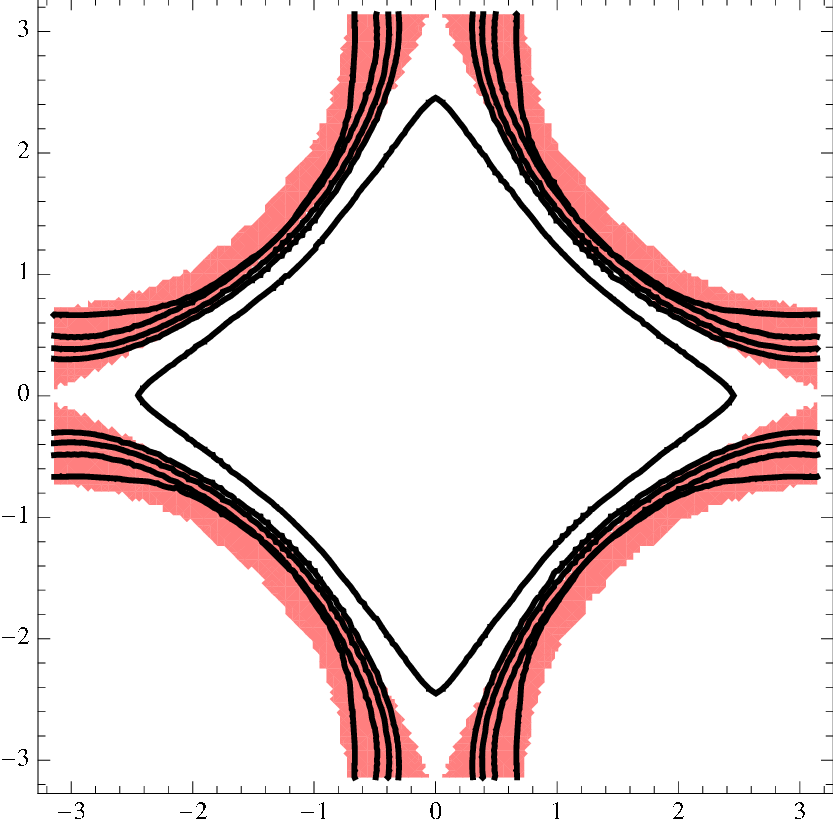} 
\end{center}
\caption{Fermi surfaces for various hole doping $h=0, 0.1, 0.2, 0.3, .04$,  as computed in section V.   The pink region (grey offline) 
is the band of attractive interactions for $g=15$.   The axes are $k_x $ and $k_y$ in the  first Brillouin zone,  i.e. in the range
$-\pi $ to $\pi$}
\vspace{ .2mm}
\label{fermisurface} 
\end{figure}

If the attractive interactions exist near the Fermi surface,  then  Cooper's original argument 
should apply:
the filled Fermi sea just serves to block states and the particles can form a bound state,
i.e. Cooper pairs\cite{Cooper}.   
Let us then make the  hypothesis  that the  regions of attractive interactions described above 
lead to Cooper pairing.     Then the  following scenario  emerges.  
The Fermi surface with interactions is calculated in section V based on the
filling fractions $f$.   
   In Figure \ref{fermisurface}  we plot these Fermi surfaces for various hole doping,
  and also display the attractive band.      (See the next section for the  precise definition of hole doping
  $h$;  as defined it corresponds to the number of holes per plaquette.)
  These computed Fermi surfaces closely parallel experimental measurements,  in 
  that they flare out in the anti-nodal directions
i  .e. $(k_x , k_y) =  (0,\pi)$ and $90^\circ$ rotations 
thereof\cite{ExpFermiSurface}.  
 This figure  shows that the attractive regions in the 
   anti-nodal directions   play the most significant role.   
At low densities (high hole-doping),   there are no attractive interactions within the
Fermi surface,  and the model should correspond to a Fermi-liquid.   On the other hand,  as 
$h$ is decreased,   
 the Fermi surface  intersects  
attractive regions in the anti-nodal directions.  
This first occurs around a hole doping $h=0.3$.    
  If a gap forms in these directions,   then this could explain the  anisotropy  of the gap,
which is zero in the nodal $(\pi, \pi)$ directions.    
As the density is increased further,  eventually the Fermi surface is beyond the attractive band. i.e. 
the attractive band is completely enclosed by  the Fermi surface.     
At half filing,  $h=0$,  the attractive regions in the anti-nodal directions are just inside the
Fermi surface,  however the Fermi surface still intersects the attractive band in the nodal,
$(\pi, \pi)$,   directions.

\section{Thermodynamics from the S-matrix}

\def\muhat{{\hat{\mu}}}

The important question that remains is whether  our approach to 
the thermodynamics of the Hubbard model
  can capture the instabilities proposed at the end of the last section based on the properties of the kernel $G$.     If so,  this  will provide  a calculation
of the critical temperature.   

In this section we describe how to compute the free energy from a formalism 
that is a synthesis of the works\cite{LeClairS,PyeTon}.       Being a synthesis, 
it is worthwhile reviewing the main features of the derivation,  and how the construction
follows from the basic ingredients in these two papers.  

The starting point is a formal expression for the partition function $Z$ in terms of the
S-matrix derived in \cite{Ma}:
\beq
\label{Zma}
Z = Z_0 + \inv{2\pi} \int dE  e^{-\beta E}  \, \Im \d_E  \log \hat{S} (E) 
\eeq
where $\hat{S} (E)$ is the off-shell S-matrix operator,  $Z_0$ the free partition function, 
and $\beta = 1/T$.  
Although the above formula is simple enough,  a considerable amount of additional work 
is needed to obtain something useful out of  it.    For instance the cluster decomposition property
of the S-matrix is needed to show that $Z$ exponentiates to an extensive free energy.  

Consider for simplicity a single species of fermions.  
The basic dynamical variables are the occupation numbers $f$ which determine the density:
\beq
\label{denf}
n = \int \frac{d^2 \kvec}{ (2\pi )^2 }  ~   f(\kvec)  
\eeq
Using a Legendre transformation in the variables $n$ and $\mu$,  where  $\mu$  is 
the chemical potential,  one can show that there exists a functional
$\digamma (f)$  such that the physical free energy 
follows from the variational principle $\delta \digamma /   \delta f   = 0$.    This functional 
can be separated into a free part $\digamma_0$ and an interacting part $\digamma_1$,
\beq
\label{Gamma}
\digamma = \digamma_0 + \digamma_1
\eeq
The interacting part contains contributions from N to N particle scattering for all N.     One expects the 2-particle  term to be the most important and is of the form: 
\beq
\label{Fone}
\digamma_1  =  -\inv{2}  \int  \frac{d^2 \kvec}{(2\pi)^2 }   \int \frac{ d^2 \kvec'}{ (2 \pi)^2}   
~  f(\kvec') \,  G(\kvec, \kvec' ) \, f(\kvec )  
\eeq
Although $G$ is built only from the 2-body S-matrix,   the formalism re-sums  all many-body processes
that involve arbitrary numbers of primitive binary collisions.    
(In \cite{LeClairS},   certain terms in (\ref{Zma}) (referred to as $Z_B$ terms)  were 
incorrectly dropped.   This was corrected in \cite{PyeTon},  which led to the expression in
the last section for the kernel $G$.)   

The primary difference of the two works \cite{LeClairS,PyeTon}  is the choice of $\digamma_0$.
The choice made in \cite{PyeTon} was better suited to the diagrammatic expansion,
and the resulting integral equation effectively sums up an infinite number of diagrams.
However it was found for the present problem that this integral equation only has solutions
in a very limited range of temperature and chemical potential,  indicating that the sum of
diagrams does not converge.    In contrast,  it turns out  the choice of $\digamma_0$  made in
\cite{LeClairS} does not suffer from this problem.     The latter $\digamma_0$ also has
an appealing physical interpretation,   as we now explain.   
Consider
\beq
\label{F0def}
\digamma_0 = \int  \frac{d^2 \kvec}{(2\pi)^2}   
\( ( \omega_\kvec - \mu  ) f   -   \inv{\beta}  \[  (f-1) \log (1-f) - f \log f \]  \)
\eeq
where $\omega_\kvec$ is the 1-particle energy of the free theory.   
The above expression can be interpreted as 
$\digamma_0 =  \ep  -  T  s $,  
where $\ep$ is the first $(\omega-\mu ) f$ term and represents the energy density.   
The remaining term represents the entropy density $s$\cite{LL}.    This choice 
of $\digamma_0$  also more closely parallels the derivation of the thermodynamic
Bethe ansatz\cite{YangYang}.  

Let us parameterize the occupation numbers in terms of a pseudo-energy $\vep$:
\beq
\label{fep}
f (\kvec )  =  \inv{ e^{\beta \vep (\kvec)} +1 } 
\eeq
Then the variational equation $\delta \digamma / \delta f= 0$  can be expressed in the simpler form:
\beq
\label{pseudointeq}
\vep (\kvec ) = t  \omega_\kvec - \mu   - t   \int \frac{ d^2 \kvec' }{(2 \pi)^2 }  ~
G(\kvec, \kvec' )  \,  \inv{  e^{\beta \vep (\kvec' )} +1 } 
\eeq
(We have restored the hopping coupling $t$ here.)  
Using the above equation in $\digamma$,  the free energy density $\CF$  can be expressed as
\beq
\label{freeEn}
\CF =  -  T  \int  \frac{d^2 \kvec}{ (2\pi)^2}   
\[   \log ( 1 + e^{-\beta \vep} )  +  \frac{\beta}{2}  \inv{ e^{\beta \vep} +1 }  ( \vep - \omega_\kvec + \mu )  \]  
\eeq
Comparing with \cite{PyeTon},  one sees that in the limit of small $G$,   the equations
presented there reduce to eqns. (\ref{pseudointeq},\ref{freeEn}).

For two-component fermions,     
  the occupation numbers  are parameterized in terms of two pseudo-energies
$\vep_{\up, \down}$,  and they satisfy a coupled system of two integral equations:
\beq
\label{twopseudo}
\vep _\up (\kvec ) = t  \omega_\kvec - \mu_\up   - t   \int \frac{ d^2 \kvec' }{(2 \pi)^2 }  ~
G(\kvec, \kvec' )  \,  \inv{  e^{\beta \vep_\down (\kvec' )} +1 } 
\eeq
and the same equation with $\up \leftrightarrow \down$.    Here the kernel $G$ is 
related to the scattering of spin up with spin down particles.   
 By the SU(2) symmetry, for equal chemical
potentials $\mu_\up = \mu_\down \equiv \mu$,   $\vep_\up = \vep_\down \equiv \vep$,  and one only needs to solve
one integral equation.   The occupation number for each spin component 
 has the form of a free theory given in (\ref{fep}),  and the total  density is 2 times the
 expression in (\ref{denf}),  as is the free energy.

 \def\That{{\hat{T}} }

  For the Hubbard model, since we scaled out $t$ and the lattice spacing, 
everything then depends on the dimensionless variables $
\muhat \equiv  \mu/t,  \That \equiv  T/t, \hat{t}'= t'/t
$.  
Note that all the temperature dependence is in $\That$,  thus possible phase transitions
should occur at fixed values of $T/t$,   for given $g, t'$.   
   Henceforth we drop the hats,
it being implicit that $T, \mu$   and $t'$   are  in units of $t$.   
The free energy density then takes the form:
\beq
\label{freeenergy}
\CF =  - \frac{T}{a^2}\,  c(\mu , T )
\eeq
where we have defined a scaling function $c$  
(we suppressed the dependence on $t'$):
\beq 
\label{ceq}
c  =  2  \int \frac{d^2 \kvec}{ (2\pi)^2} \[ 
  \log \( 1 +   e^{-\beta  \vep (\kvec  )   }  \) 
+ \frac{\beta}{2}  \inv{e^{\beta \vep } + 1 }  (  \vep(\kvec) -  \omega_\kvec + \mu  \, )    \]  
\eeq
It will also be convenient to express the density $n=-\d \CF / \d \mu$ as
\beq
\label{denz}
n  =   \frac{2q(\mu, T ) }{a^2} = \frac{1-h}{a^2}  
\eeq
where 
\beq
\label{qdef}
q  = 
\int \frac{d^2 \kvec}{(2\pi)^2}  \,  \inv{ 
e^{\beta \vep (\kvec)} + 1}
\eeq
Since $2q$ is the number of particles of either spin per lattice site, 
 half-filling corresponds
to $q=1/2$.   The quantity $h$ then corresponds to hole doping  when 
it is  positive,  otherwise it represents particle  doping.   More precisely,  $h$ is the number of
holes per plaquette and the lattice is completely depopulated at $h=1$.

In order to probe the properties of the model,  we will need a few 
other thermodynamic quantities.  As usual the pressure $p= -\CF$. 
 Consider first the entropy  per particle,
$S/N =  s/n$, where the entropy density $s= - \d \CF / \d T$.  It
can be expressed in terms of the scaling functions as follows:
\beq
\label{sn} 
\frac{S}{N}=     \inv{2q} \( c  + T \d_T  c \)    
\eeq
The energy density $\ep =  E/V =  T s + \mu n  + \CF $.
Thus the  energy per particle is
\beq
\label{enperpart}
\frac{E}{N t } =  \frac{\ep}{n t } = \mu  +  \frac{T^2 }{2q}  c 
\eeq

The specific heat per particle $C_V/N$ at constant volume and 
particle number $N$ is slightly more complicated since one must impose
the constant density constraint.   Setting $d q/ dT =0$ relates 
$\mu $ and $T$ derivatives as follows:
\beq
\label{dtdx}
\frac{\d T }{\d \mu}  =  -   \frac{\d_\mu q}{\d_T q} 
\eeq
Using this,  the specific heat per particle has the following expression:
\beq
\label{CVN}
\frac{C_V}{N} =  \inv{N} \( \frac{ \d E}{\d T} \)_{N,V} 
=   
\frac{T}{q} \(   \d_T  c     + \frac{T}{2} \d_T^2  c   
-    \frac{ (\d_T q)^2 }{\d_\mu q }   \)  
\eeq

\section{The free energy and estimates of critical temperatures}

In this section we analyze the thermodynamics based on the formulas of the last section, 
provide evidence for instabilities,  which may perhaps be phase transitions,  and estimate  critical temperatures.   

In order to study  the free energy,  one must first solve the integral equation 
(\ref{pseudointeq})  for the pseudo-energy $\vep$.    This can be done iteratively,
i.e. one starts with the approximation $\vep_0 = \omega_\kvec  - \mu$ and plugs this into the
right hand side to generate $\vep_1$;  this procedure is repeated until the solution converges.     
We approximated the integral equation by 
approximating the Brillouin zone as a $10\times 10$ grid,   and performing the integrals as discrete sums.   
This is rather crude,  and was  due to our limited computing resources;  certainly one can do better.  
It was found  that  for  large portions of the $\mu, T$ parameter space,   the iterative procedure
converged rapidly,   typically within 10 iterations.        

\begin{figure}[htb] 
\begin{center}
\hspace{-15mm} 
\psfrag{x}{$\mu$}
\psfrag{y}{$h$}
\includegraphics[width=10cm]{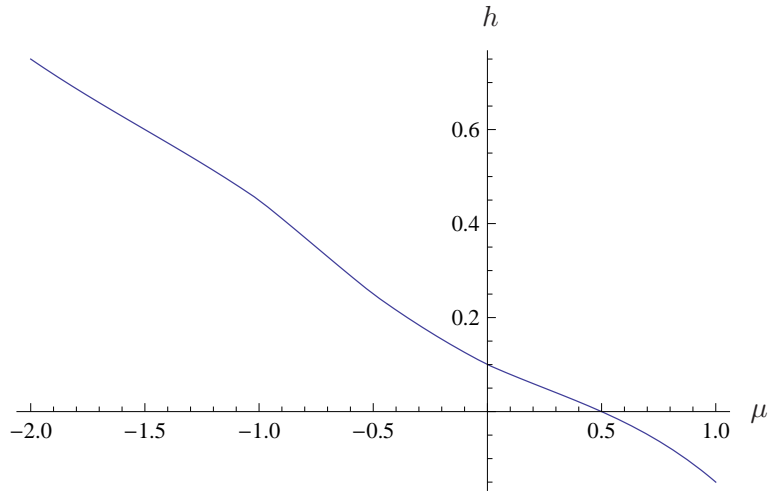} 
\end{center}
\caption{Hole doping $h$ as a function of chemical potential for $T=0.2$.}  
\vspace{-2mm}
\label{hofmu} 
\end{figure}

For reasons  stated above,  our analysis was performed for $g =15$,
  and $t'/t = -0.3$.  For fixed doping $h$,  the chemical potential depends on
  temperature,  but we find this dependence to be weak.   
In Figure \ref{hofmu} we plot hole doping $h$ as a function of chemical potential 
at the low temperature $T=0.2$,  and it is nearly a straight line.   
  As expected,   increased doping corresponds to 
decreasing chemical potential;  half filling occurs around $\mu = 0.5$.       
This positive value of $\mu$ at half-filling is due to the mainly repulsive interactions.

The computed Fermi surfaces for various hole doping 
 are shown in Figure \ref{fermisurface}.   In this figure they
are defined as the contour where the filling fraction $f= 1/2$  at  the low   temperature $T=0.025$.  
They are in good agreement with experiments\cite{ExpFermiSurface},  especially in the anti-nodal directions,   whereas in the
nodal directions  they are pulled back toward the center in a more pronounced manner in the data. 
This can likely be accounted for by adding additional hopping terms,  such as  next-to-next nearest neighbor.

\begin{figure}[htb] 
\begin{center}
\hspace{-15mm} 
\includegraphics[width=10cm]{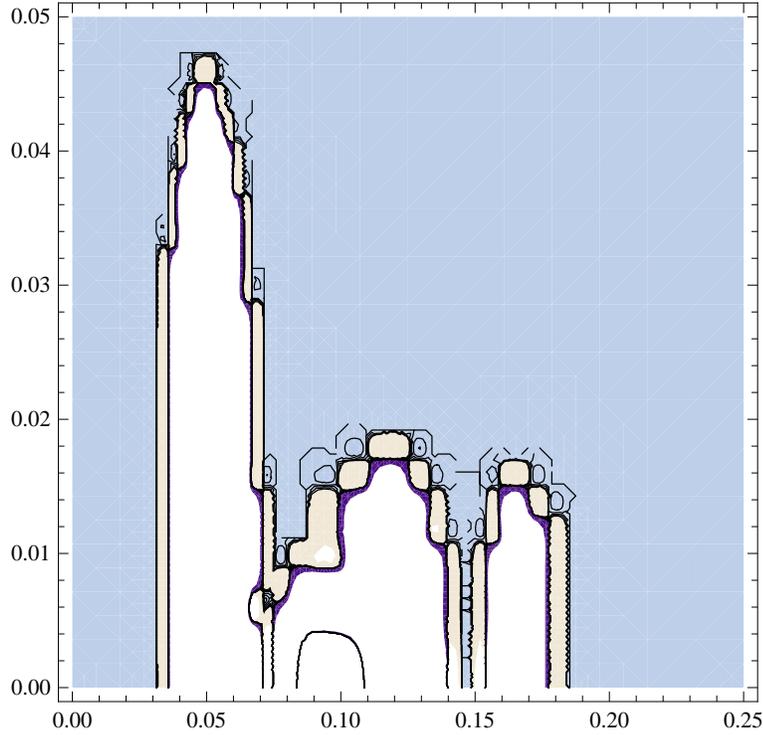} 
\end{center}
\caption{Existence of solutions based on the iterative method. 
  In the light   regions there are no solutions to the 
integral equation for the pseudo-energy.  The horizontal axis is the 
hole doping $h$,  and the vertical axis is the temperature $T$.
(See text for a detailed explanation.) }  
\vspace{-2mm}
\label{yesInt} 
\end{figure}

The most interesting feature of the integral equation (\ref{pseudointeq}) is that
there are regions in $\mu, T$ where  the iterative procedure does not lead to a solution
for an arbitarily high number of iterations;  the procedure leads to $\vep$ that jumps successively
between two values,  neither of which are solutions.  (Figure \ref{yesInt}.)  Let us assume that in these regions,  no 
solution exists,  although our observations do not necessarily prove this.   
Furthermore,  let us adopt the following  physical  interpretation.    By comparison,  in the standard BCS theory
of superconductivity,   one has  a finite temperature gap equation.   As the temperature 
is raised,   at the critical temperature there are no longer  solutions to this gap equation;  i.e. 
as far temperature is concerned,  it is a bottom up approach.   In contrast,  the present formalism
is a top down approach:   as the temperature is lowered one reaches a critical temperature 
where solutions no longer exist.    Let us  interpret this as an instability toward formation of a new phase,
or perhaps a cross-over to a different behaviour.
In support of this interpretation,  we mention the treatment of the unitary Bose gas within the
present formalism\cite{PyeTon2}.   The gas undergoes a phase transition  to a Bose-Einstein condensate at a critical value
of $\mu/T$;  above this value there are no solutions to the pseudo-energy  integral equation.  

  It should be emphasized that  the  true nature of this `phase'  cannot be surmised from our thermodynamic approach alone;   in 
addition one needs a bottom up approach that contains information about the zero temperature ground state,  such as a gap equation.      Such a complementary bottom up approach is
developed in \cite{HubbardGap},  where solutions to a BCS-like  gap equation 
based on the attractive interactions described in section III are studied.    
The solutions are highly anisotropic,  in that they vanish in the nodal directions,  and are largest in the anti-nodal, 
 and the critical $T_c
\approx 0.04$ found there for $h=0.15$  is  consistent with the critical temperatures estimated below.

  Figure \ref{yesInt}  indicates  the regions where solutions do not exist for positive
hole doping $h<0.25$.   This figure is a contour plot of an interpolating function 
defined to be  equal to 1 if there is a solution,  and zero otherwise;   the white region 
indicates the region of no  solution, whereas in the light blue region, solutions exist.   
The  boundary between the regions of existence and non-existence of solutions are the darkest curves, 
which are reasonably well delineated.
Due to the 2-body approximation we have made in the thermodynamics,  this boundary is not to
be taken as precisely determined.   
The roughness of the boundaries we believe is a numerical artifact,
mainly attributed to not  using a fine enough grid in the temperature and chemical potential variables. 
The dip around $h=0.15$ we also believe to be an  artifact since it disappears upon varying $g$
and $t'$.     
In the range of doping $0.03 < h < 0.2$ one sees a possible phase transition with critical temperatures
ranging from  $0<T_c< .05$.    
  As explained in Section III, since this is the range of doping where the Fermi surface is intersecting the attractive band in the anti-nodal directions,   we propose that this signifies an instability toward the formation of Cooper pairs, 
 so that superconductivity may occur in the 
white regions.  
At hole doping $h=0.15$,  $T_c \approx 0.02$.  
This is reasonable,  since experimentally  $T_{c, {\rm max}} /t  \approx 0.025$.

\begin{figure}[htb] 
\begin{center}
\hspace{-15mm} 
\psfrag{x}{$T/t$}
\psfrag{y}{$\vep (0, \pi) + \mu$}
\includegraphics[width=10cm]{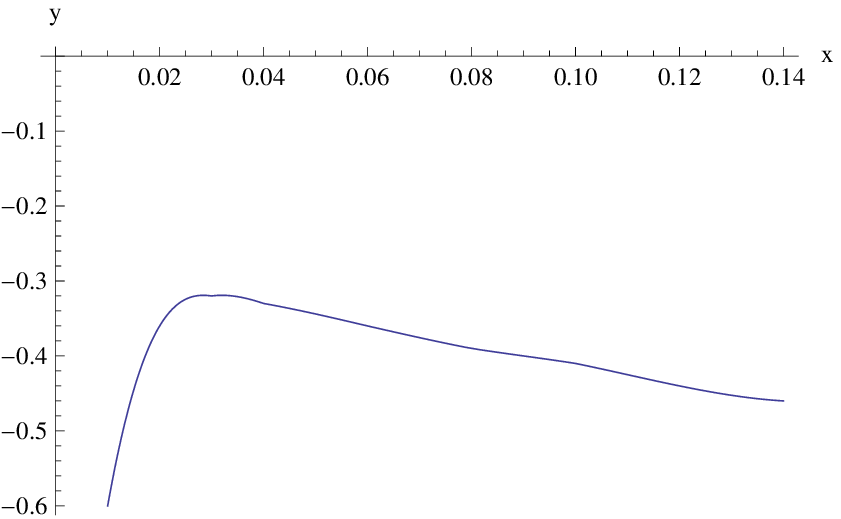} 
\end{center}
\caption{The single particle energy $\vep (\kvec ) + \mu$  in the anti-nodal direction 
$\kvec = (0, \pi)$  at optimal hole doping $h=0.15$  as a function of temperature. }  
\vspace{-2mm}
\label{vep.15} 
\end{figure}

The single quasi-particle energies correspond to  $\vep (\kvec )  + \mu$.    
In Figure \ref{vep.15}  we plot this single particle energy as a function of temperature 
at optimal hole doping $h=0.15$ in the anti-nodal direction.   One clearly sees a drop 
at $T_{c, {\rm max}}$.

The higher $T_c$'s up to $0.05$ in the strongly underdoped region possibly 
signify the so-called  pseudogap scale $T^*$.   There appears to be a small separation
around $h=.08$,  however this is less pronounced for other $g$,  so it is not
clear if this signifies anything.     
Some recent experimental results are very relevant to the issue \cite{Seamus,Chatterjee}.  
Remarkably, it 
was found that the  superconducting gap smoothly evolves into 
the  pseudogap,  i.e. they both seem to arise from the same underlying
mechanism.   
In other words the  gap is physically present even in regions of no 
superconductivity.   This suggests that the $T_c$'s  in Figure \ref{yesInt} 
may all be arising from the same underlying phenomenon.    
This is  consistent with the complimentary gap equation analysis in \cite{HubbardGap}, 
where it was found that the gap extends and increases  all the way to zero doping,  
as does the critical temperature scale  in Figure \ref{yesInt}.    
Although not shown in Figure \ref{yesInt},   at higher doping there
is another region of no-solutions with a maximum $T/t \approx 0.07$.   This  could perhaps signify
the temperature referred to as  $T_{\rm coh}$ in the literature,   where a crossover in the 
resistivity is observed from $\rho \propto T$  to $\rho \propto T+ T^2$\cite{Hussey}.

\begin{figure}[htb] 
\begin{center}
\hspace{-15mm} 
\includegraphics[width=10cm]{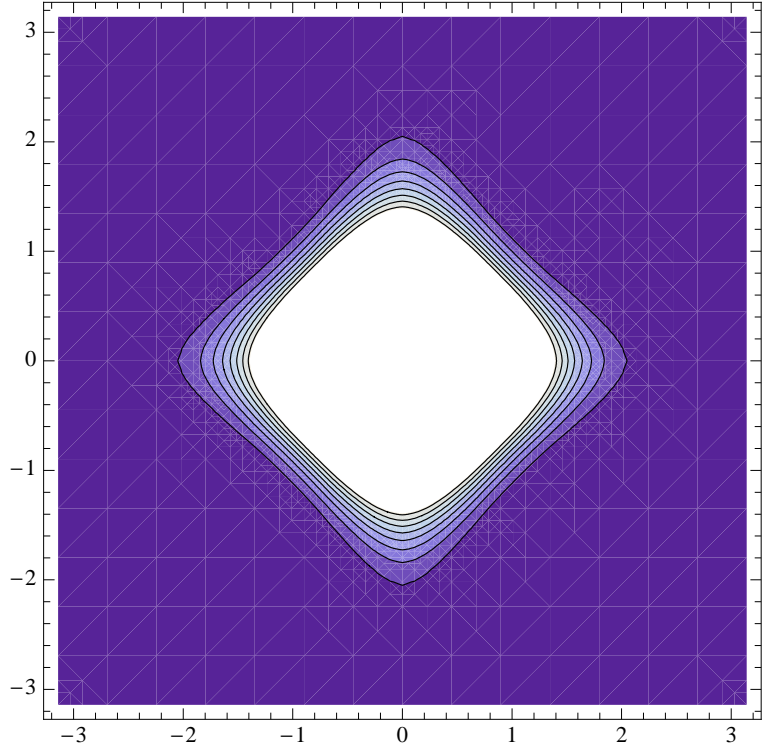} 
\end{center}
\caption{The Fermi surface  for hole doping $h=0.8$ at temperature $T=0.2$.
  The white region 
has $f=1$ and the darkest (purple) region $f=0$.}  
\vspace{-2mm}
\label{FS.8} 
\end{figure}

\begin{figure}[htb] 
\begin{center}
\hspace{-15mm} 
\psfrag{x}{$T$}
\psfrag{y}{$E/N$}
\includegraphics[width=10cm]{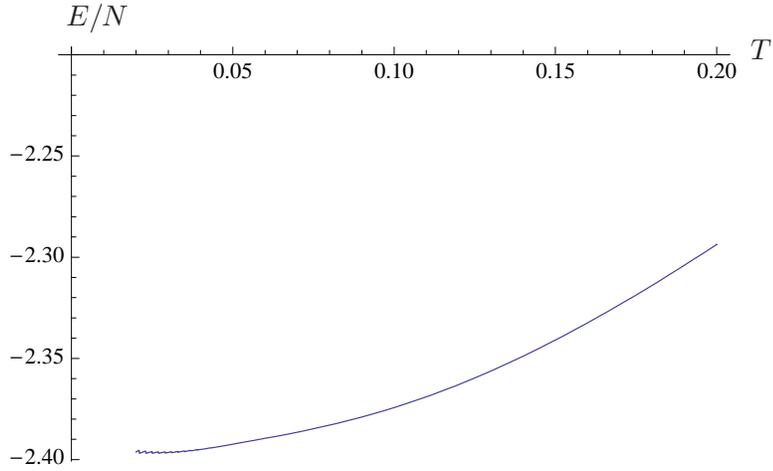} 
\end{center}
\caption{The energy per particle as a function of temperature for hole doping $h=0.8$.}  
\vspace{-2mm}
\label{energy} 
\end{figure}

\begin{figure}[htb] 
\begin{center}
\hspace{-15mm} 
\psfrag{x}{$T$}
\psfrag{y}{$S/N$}
\includegraphics[width=10cm]{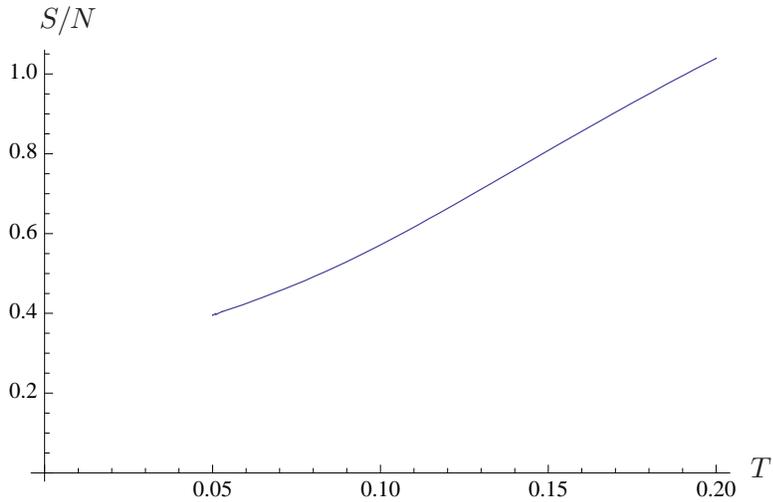} 
\end{center}
\caption{Entropy per  particle as a function of temperature for hole doping $h=0.8$.}  
\vspace{-2mm}
\label{entropy} 
\end{figure}

\begin{figure}[htb] 
\begin{center}
\hspace{-15mm} 
\psfrag{x}{$T$}
\psfrag{y}{$C_V / N $}
\includegraphics[width=10cm]{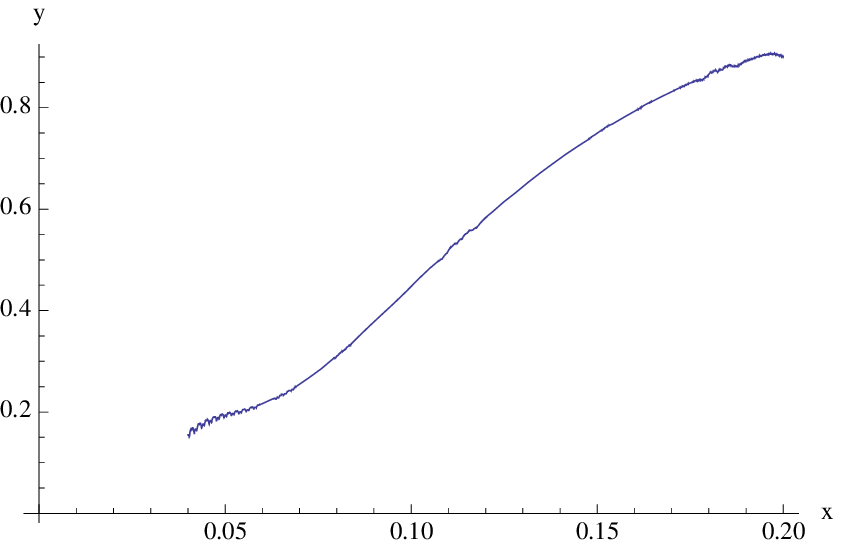} 
\end{center}
\caption{ Specific heat per particle as a function of temperature for hole doping $h=0.8$.}  
\vspace{-2mm}
\label{specificheat} 
\end{figure}

We turn next to the thermodynamic functions,   such as energy and entropy per particle.  
For hole densities in the vicinity of the boundaries shown in Figure \ref{yesInt}, 
our  crude  solution  to the integral equation for the pseudo-energy  
is not smooth enough to reliably 
compute temperature derivatives numerically.  
However,  at low density our numerical results are better behaved,  
and although of less  interest physically for the cuprates,   at least allow a comparison with
previous literature.   
  We  therefore
analyzed  the thermodynamics in the overdoped region,  with $h=0.8$.  
The Fermi surface is shown in Figure \ref{FS.8}.
In Figures \ref{energy},   \ref{entropy}  and \ref{specificheat}, 
we plot the energy and entropy per particle and specific heat  as a function of 
temperature.       Our results for the entropy and specific heat  
are roughly consistent with the results in \cite{Bonca,Li},  especially the results in 
\cite{Bonca},  which extend to low density;  a detailed comparison is beyond reach since
previous results are typically at higher temperatures.

\vfill\eject

\section{Conclusions}

We  presented  an  analytic treatment of the two-dimensional Hubbard model   at finite
chemical potential and temperature 
based on  a new approach to statistical mechanics we recently developed\cite{LeClairS,PyeTon},
which is built upon  the S-matrix.    The effective momentum-dependent coupling in this approach is
the kernel $G$ of an integral equation satisfied by the pseudo-energy $\vep$,  which is built on the
exact two-body S-matrix.     
We showed that there are regions of the Brillouin zone where the interactions are effectively
attractive, for example,  $U/t > 6.4$  for $t'/t = -0.3$, 
  even though the bare model only has repulsive interactions,  and this is essentially 
due to multi-loop quantum corrections.    The next-to-nearest neighbor
hopping coupling $t'$  plays a significant role in determining this property,
  and  our analysis suggests that  for a fixed value of $U/t$,  superconductivity may not exist for $t'=0$. 
We emphasize that no approximations were made in obtaining these results;   e.g. 
 the existence of a  threshold in $U/t$ for the existence of certain attractive interactions,
  stands on its own and is  independent of the subsequent approximations we made in
the thermodynamics.      

We postulated that phase transitions occur where there are no solutions of the integral equation
for the pseudo-energy.       
On the overdoped side,  this phase sets in  at hole doping $h<0.25$.    
Our result for $T_c  \approx 0.02$  at  $h=0.15$ is in good agreement with experiments.   
We found that there is also evidence for transitions in the underdoped region,  and we  suggested
this  signifies the pseudogap.

The main lesson of this work is that quantum loop corrections to scattering  are perhaps the origin of 
the attractive interactions that lead to Cooper pairing near the Fermi surface.   If this idea is correct, 
then in order to complete the picture one needs to derive a gap-equation that describes the
structure of the ground state at zero temperature.    Some preliminary attempts in this direction 
were taken  in \cite{HubbardGap},  
where solutions to a gap equation based on the attractive interactions described above are studied,
and the critical temperatures found are consistent with the bottom down approach described in
this paper.  Based on the  detailed properties of the solutions to the gap equation studied  there,  i.e. its anisotropy
and the existence of  Fermi arcs in the nodal direction,  it was suggested that the attractive mechanism in
this paper may be responsible for the pseudogap rather than d-wave superconductivity.

\section{Acknowledgments}

   I would like to thank Henry Tye for discussions and  Eliot Kapit  and Kyle Shen  for their 
    help in  understanding 
    the phenomenology of high $T_c$ superconductivity.   
   This work is supported by the National Science Foundation
under grant number  NSF-PHY-0757868.

\end{document}